\documentclass[twocolumn,showpacs,preprintnumbers]{revtex4}%
\usepackage{graphicx}
\usepackage{dcolumn}
\usepackage{bm}
\usepackage[latin1]{inputenc}
\usepackage{amsmath}%
\usepackage{amsfonts}%
\usepackage{amssymb}

\begin{document}

\title{Optomechanical damping of a nanomembrane inside an optical ring cavity}
\author{Arzu Yilmaz, Simon Schuster, Philip Wolf, Dag Schmidt, Max Eisele, Claus
Zimmermann, Sebastian Slama}
\affiliation{Physikalisches Institut and Center for Collective Quantum Phenomena in LISA+, Eberhard Karls Universit\"{a}t T\"ubingen, Auf der Morgenstelle 14, D-72076 T\"ubingen, Germany}
\affiliation{}
\date{\today}

\begin{abstract}
We experimentally and theoretically investigate mechanical nanooscillators
coupled to the light in an optical ring resonator made of dielectric mirrors.
We identify an optomechanical damping mechanism that is fundamentally
different to the well known cooling in standing wave cavities. While, in a
standing wave cavity the mechanical oscillation shifts the resonance frequency
of the cavity in a ring resonator the frequency does not change. Instead the
position of the nodes is shifted with the mechanical excursion. We derive the
damping rates and test the results experimentally with a silicon-nitride
nanomembrane. It turns out that scattering from small imperfections of the
dielectric mirror coatings has to be taken into account to explain the value
of the measured damping rate. We extend our theoretical model and regard a
second reflector in the cavity that captures the effects of mirror back
scattering. This model can be used to also describe the situation of two
membranes that both interact with the cavity fields. This may be interesting
for future work on synchronization of distant oscillators that are coupled by
intracavity light fields.

\end{abstract}
\maketitle

\section{Introduction}

Optomechanical forces acting on micro- and nanomechanical systems have been
intensively investigated due to their fundamental aspects in quantum mechanics
of macroscopic bodies and because of possible applications in quantum
metrology and hybrid quantum systems \cite{Aspelmeyer14}. Important progress
has been achieved during the last years in cooling mechanical oscillators to
their vibrational ground state \cite{Teufel11, Chan11}.
The underlying damping mechanism involves the coupling of a mechanical mode of
the oscillator with an optical mode that is typically provided by the light
field inside a standing wave cavity. Complementary setups have been realized:
cavities with an oscillating end mirror \cite{Metzger08, Usami12},
membrane-in-the-middle (MIM) cavities \cite{Thompson08, Wilson09}, and
evanescent-wave resonators \cite{Anetsberger09}. In all of these situations
the optomechanical interaction is based on the shift of the cavity mode
resonance frequency $\omega_{c}(x)$ when the mechanical oscillator is
displaced by $x$. This is quantified in the parameter $G=\frac{d\omega_{c}%
}{dx}$ \cite{Aspelmeyer14}.

\begin{figure*}[th]
\centerline{\scalebox{.8}{\includegraphics{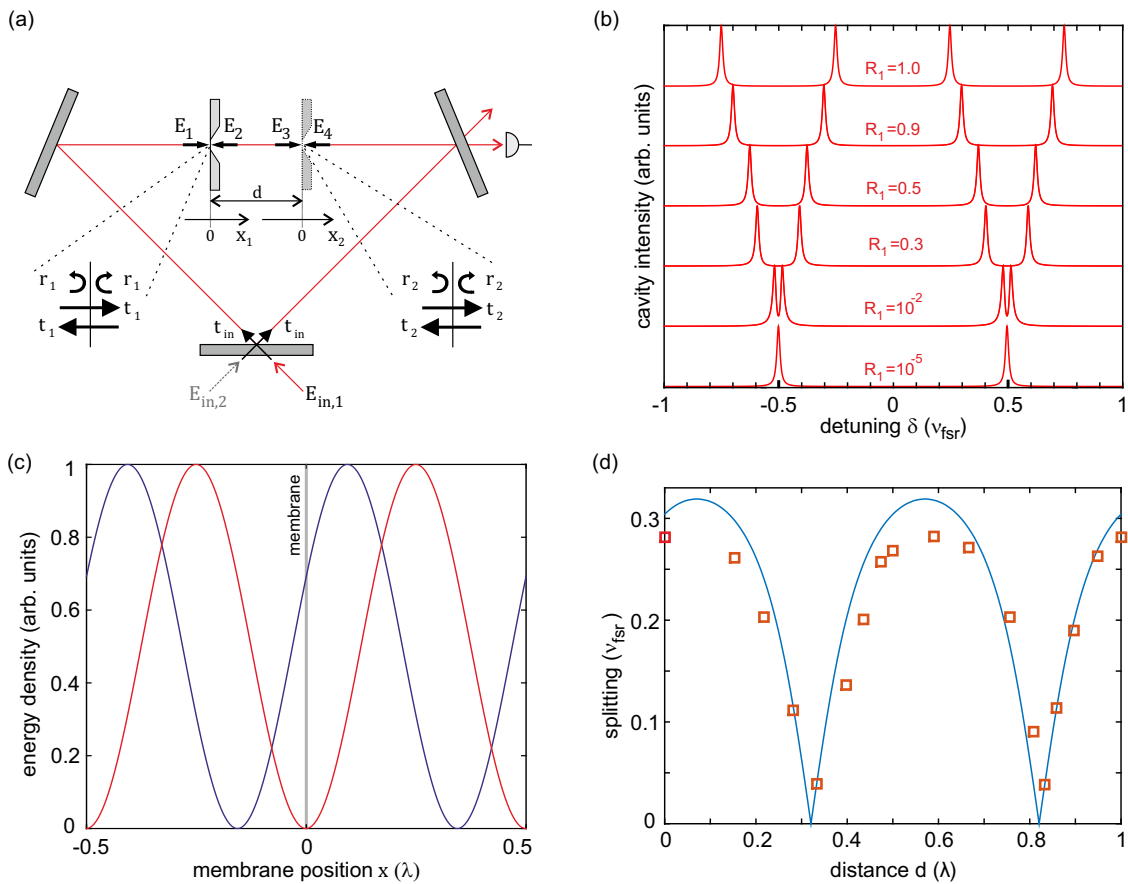}}}\caption{(a) Sketch of
the investigated setup. A ring resonator made from high quality dielectric
mirrors contains two nano membranes with amplitude reflection $r_{i}$,
transmission $t_{i}$ and time dependent excursion $x_{i}$. The fields $E_{1}$
to $E_{4}$ are each evaluated at an infinitesimal displacement to the left
resp. to the right of the membranes equilibrium position and are propagating
clock-wise resp. counter clock-wise in the cavity. (b) Calculated spectra of a
ring resonator that contains a single membrane with various reflection
coefficient $R_{1}=\left|  r_{1}\right|  ^{2}$. The reflectivity of the second
membrane is set to zero, $R_{2}=0$. The membrane couples the two circulating
waves and lifts the mode degeneracy of an empty cavity. (c) Calculated energy
density distribution of the standing light waves in the cavity for both eigenmodes with $R_{1}=0.3$ and $R_{2}=0$. (d) Calculated mode splitting (blue
curve) for a ring cavity with two membranes with various separations
($R_{1}=R_{2}=0.3$). Red squares denote the experimentally observed splitting
with the cavity in air.}%
\label{fig:setup}%
\end{figure*}For a membrane in a high finesse optical ring resonator made from
standard dielectric mirrors the optomechanical damping is fundamentally
different. An empty ring cavity supports two degenerate modes formed by two
counter circulating travelling waves. As long as there is no coupling between
the waves their relative phases are independent. A membrane inside the cavity
scatters light between the two modes and locks their phases. This mode
coupling also lifts the degeneracy which leads to a line splitting in the
cavity spectrum. The two new eigenmodes are standing waves formed by
orthogonal superpositions of the circulating waves. The crucial point now is
that the nodes of these standing waves are locked to the position of the
membrane. If the membrane is displaced in the cavity the standing waves are
displaced by the same amount without changing their eigenfrequencies. Unlike
in linear cavities, the steady-state resonance frequency in an ideal ring
cavity does not depend on the position of the membrane, i.e. $G\equiv0$. The
standard theory for optomechanical damping would thus predict no damping at
all. This is probably the reason why such ring cavities have not been
discussed so far in the context of damping the motion of nanomechanical
oscillators. \cite{Sawadsky15_footnote, Sawadsky15}. What has been studied
extensively is the coupling of ring cavity modes to the mechanical motion of
cold atom clouds consisting of many thousands of atomic oscillators with
typically much smaller reflectivity than that of a dielectric membrane
\cite{Ritsch13}. The rich physics of such coupled atom-photon-systems includes
nonlinear dynamics, optical bistability, strong coupling, collective
scattering, dynamic instabilities, and self-synchronization \cite{Nagorny03,
Elsaesser04, Klinner06, Kruse03, Slama07,Schmidt14, Cube04}, topics which have
all been discussed also in the context of cavity optomechanics
\cite{Aspelmeyer14}. Moreover, Gangl et al. predicted that cooling of
particles can be even more efficient in ring cavities than in standing wave
cavities \cite{Gangl00}. It thus seems promising to start experiments with
macroscopic oscillators inside mirror ring resonators.

As it turned out, the below presented observed damping rates cannot be
understood without taking into account mirror back scattering. Residual
imperfections of the dielectric mirror coatings can rescatter light between
the two circulating modes of the ring resonator. One way to theoretically
describe such back scattering is to introduce a fictitious partially
reflecting intra cavity mirror that effectively captures the total scattering
at all optical surfaces inside the resonator. Such a model, however, can
easily be extended to describe two membranes in an ring resonator with the
fictitious mirror now realized by the second membrane. With this in mind, we
start Section II with deriving the equations for a ring cavity that contains
two membranes. We calculate the steady state spectra and experimentally test
them in air. Section III describes the main experiment and presents the
observed mechanical damping rates in a single membrane resonator in vacuum. In
the theoretical Section IV we write down the equations of motion and introduce
a pertubation method to derive the optomechanical damping rates. We first
apply the model to a single membrane in an ideal resonator. It can be solved
fully analytically and is less involved than the full model for two membranes.
In a second step, backscattering is included and the damping rate for the
first membrane is calculated semi analytically with the second membrane held
fixed in order to represent back scattering. Finally, the result is discussed
and compared to damping in linear cavities. Section V concludes with an
outlook to the physics in real ring resonators with two membranes.

\section{Resonator spectra with one and two membranes}

The scenario discussed here is shown in Fig.~\ref{fig:setup} (a). A three
mirror ring resonator contains two membranes with complex amplitude
reflectivity $r_{1}$, $r_{2}$ and transmittivity $t_{1}$, $t_{2}$. The
membranes are at equilibrium positions $d_{1}$ and $d_{2}$ measured from the
input coupling mirror clock-wise and counter clock-wise, respectively. They
can oscillate around this position with a time dependent excursion $x_{i}(t)$.
The cavity round-trip length $l$ sets the value of the free spectral range
$\nu_{fsr}=c/l$ with $c$ being the vacuum speed of light. The distance between
the equilibrium positions of the membranes is $d=l-d_{1}-d_{2}$. We describe
the light inside the resonator in the basis of four travelling waves. Their
steady state electric field amplitudes immediately left and right of the two
membranes can be found by requiring self consistency after one round trip time
$\tau=1/\nu_{fsr}$:
\begin{align}
E_{1}  &  =t_{2}E_{3}r_{c}e^{-i\varphi_{d}}+r_{2}E_{4}r_{c}e^{-i\varphi
_{d}-2ikx_{2}}+t_{in}E_{in,1}e^{i\varphi_{1}}\nonumber\label{eq:statfields}\\
E_{2}  &  =r_{2}E_{3}e^{i\varphi_{d}+2ikx_{2}}+t_{2}E_{4}e^{i\varphi_{d}%
}\nonumber\\
E_{3}  &  =r_{1}E_{2}e^{i\varphi_{d}-2ikx_{1}}+t_{1}E_{1}e^{i\varphi_{d}%
}\nonumber\\
E_{4}  &  =t_{1}E_{2}r_{c}e^{-i\varphi_{d}}+r_{1}E_{1}r_{c}e^{-i\varphi
_{d}+2ikx_{1}}+t_{in}E_{in,2}e^{i\varphi_{2}}.\nonumber\\
&
\end{align}
Here, $r_{c}=r_{l}e^{ikl}$, $\varphi_{d}=kd$, $\varphi_{1}=kd_{1}$, and
$\varphi_{2}=kd_{2}$. The real valued parameter $r_{l}$ describes the field
amplitude reduction after one cavity round-trip due to transmission at the
input coupling mirror. Uncontrolled losses due to scattering and absorption at
the optical elements are also absorbed in an effective value for $r_{l}$. In
equilibrium the excursions vanish, $x_{i}\left(  t\right)  =0$. We solve Eq.
(\ref{eq:statfields}) numerically for unidirectional pumping, i.e.
$E_{\mathrm{in,2}}=0$, and plot the intra cavity laser intensities as a
function of detuning. Spectra for a single membrane in an ideal resonator
($R_{2}=0$) are shown in Fig.~\ref{fig:setup}~(b). The coupling of the
counterpropagating waves by the membrane leads to a splitting of the two
eigenmodes that increases with $R_{1}$. In the limit of $R_{1}=1$ the ring
transforms into a standing wave cavity and the resulting equidistant spectrum
is that of a linear cavity with a free spectral range which is half of that of
the ring cavity. In the experimentally observed spectra (not shown here) the
resonant intensity of the two eigenmodes is not exactly equal. This can be
explained by the light intensity at the position of the membrane which is
different for the two modes, see Fig.~\ref{fig:setup}~(c). The amount of
absorption inside the membrane is thus different. The plotted light
distribution is particular for modeling the membrane as an infinitesimally
thin dipole sheet which results in a specific phase of its complex
reflectivity $r_{1}$. For comparison, we have also modeled the membrane as an
extended layer with 50 nm thickness and calculated its total field
reflectivity from the single boundary Fresnel reflectivities at both sides of
the membrane. Using this more realistic model the position of the standing
wave node is slightly shifted from the membrane position, but the qualitative
behavior stays the same.

If the position of a single membrane is shifted, the resonance frequencies of
the cavity modes do not change, i.e. for a single membrane in an ideal
resonator $G=0$. This is different if the cavity contains a second membrane.
It generates its own standing wave in the cavity that adds to the standing
wave generated by the first membrane. With increasing $R_{2}$ the phase of the
total standing wave decouples from the position of the first membrane and $G$
can now differ from zero. The two membranes form an intra cavity Fabry-Perot
etalon whose reflectivity depends on the membrane separation $d$. For
destructive interference in reflection, the cavity spectrum is that of an
ideal ring cavity without any internal reflector, i.e. the splitting observed
in Fig.~\ref{fig:setup} (b) disappears. For constructive interference in
reflection, the overall reflectivity of the intra cavity Fabry-Perot etalon
becomes maximum and the resulting splitting exceeds that of an ideal resonator
with a single membrane. This behavior is captured in our model and we can
compare the simulated splitting for $R_{1}=R_{2}=0.3$ to our experimental
observation with two membranes in the ring cavity (Fig.~\ref{fig:setup}~(d).
The experiment is performed in air and we vary the distance $d$ with a
piezo-transducer to which one of the membranes was attached. Note that the
signal is periodic in $\lambda/2$. The plotted distance is thus defined only
modulo half an optical wavelength $\lambda$. Simulations and measured data
agree reasonably, with a global offset of the experimental value of $d$ as
only free parameter. From this offset all other distances are determined from
the piezo voltage. We have also tested the model by recording the splitting
caused by a single membrane, again with good agreement. Details about the
resonator geometry are found in the next section.

\section{Experimental Observations}

\begin{figure}[th]
\centerline{\scalebox{.9}{\includegraphics{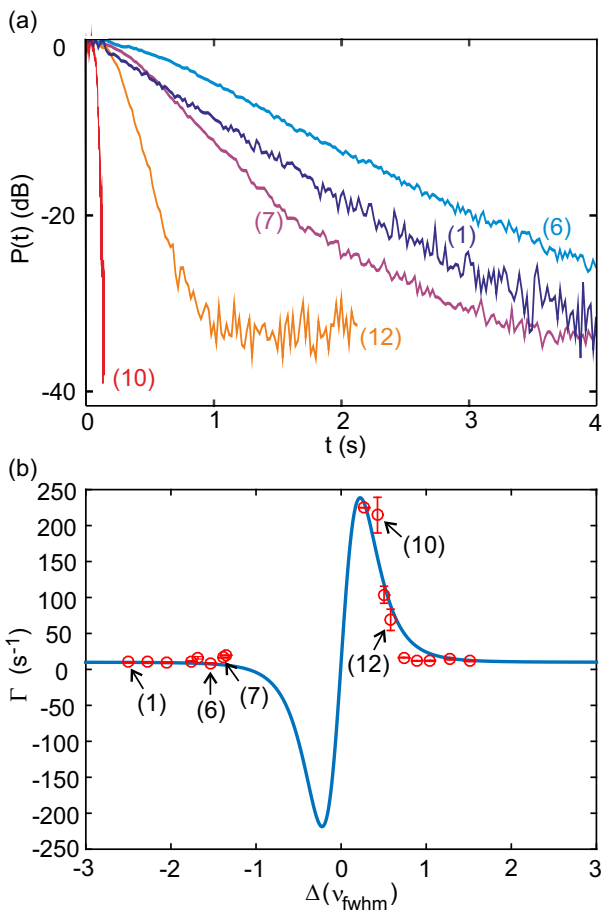}}}\caption{(a) Ring-down
measurements of membrane oscillation. For clarity, only some of the curves are
plotted, the number indicating the corresponding data point in the part (b) of
the figure. (b) Experimentally determined damping rates. The error bars
indicate the statistical error by evaluating several curves with identical
detuning. The solid line is plotted solely for guiding the eye and
illustrating the dispersive lineshape of the data points.}%
\label{fig:damping}%
\end{figure}Experimentally it is difficult to keep a membrane perfectly
aligned over a longer time in vacuum. In standing wave cavities a compact
setup helps. For a ring resonator this approach has its limits and it is not
yet clear how to solve the alignment problem with even two membranes in the
resonator. We thus restrict ourselves to experiments with a single membrane.
We glued the membrane (Norcada NX5100A, low-stress, $1~\mathrm{mm}^{2}%
\times50~\mathrm{nm}$) on a piezo-transducer and positioned it in the mode
volume of the ring cavity with round-trip length $l=8.5~\mathrm{cm}$. The
angle of the membrane with respect to the cavity axis is adjusted for normal
incidence by means of a compact and stable precision mirror mount. The finesse
of the cavity including the membrane has been determined to be $F=520$ by
modulating the phase of the laser beam with a frequency of $20$ $\mathrm{MHz}$
and using the resulting side bands as frequency markers for measuring the linewidth of the resonator. The cavity mirrors are dielectric ''super mirrors''
with specified scattering losses below $5$ $\mathrm{ppm}$. These extreme
values are only reached in a clean room environment and may be somewhat higher
in our experiment. The transmission of the input coupler is $T_{in}=\left|
t_{in}\right|  ^{2}=6.04\times10^{-3}$. The cavity including the nanomembrane
is placed in a vacuum chamber at a pressure of $p=2\times10^{-8}%
~\mathrm{mbar}$ and is single sided pumped with $P_{in}=10~\mathrm{mW}$ by a
standard grating stabilized diode laser at a wavelength near $780\;\mathrm{nm}%
$. The laser frequency $\nu$ is electronically stabilized to the resonance
frequency of the cavity $\nu_{c}$ with variable detuning $\Delta=\nu_{c}-\nu$.
Once the laser frequency is locked, the membrane is mechanically excited with
a piezo-transducer that is harmonically driven at the resonance frequency of
the membrane $\Omega_{m}\simeq2\pi\times130\,\mathrm{kHz}$. The oscillation of
the membrane is detected by overlapping the two counter-propagating modes on a
beam splitter outside the cavity. Due to the motion of the membrane the
electric fields of the two modes are phase-modulated with the membrane
frequency $\Omega_{m}$. The detected beat signal at $\Omega_{m}$ is analyzed
with a spectrum analyzer in zero-span mode. The measured power signal $P(t)$
is thus proportional to the quadratic displacement $x^{2}(t)$ of the membrane.
After exciting the membrane, we quickly switch-off the piezo drive and measure
the decrease of $P(t)$ in a ring-down experiment, see Fig.~\ref{fig:damping}
(a). We determine the damping rate $\Gamma$ by an exponential fit
$P(t)\propto\exp\left(  -2\Gamma t\right)  $. The result is plotted in
Fig.~\ref{fig:damping} (b). No data could be taken close to resonance for
negative detuning because of strong self-sustained oscillations. For positive
detuning the maximum damping rate amounts to about $250\;\,\mathrm{s}^{-1}$.
The intrinsic damping of the membrane was determined at large detuning to be
$10~\mathrm{s}^{-1}$.

\section{Optomechanical damping of a single membrane in a ring cavity}

In an ideal ring resonator without mirror back scattering a single membrane is
the only element that couples the two counter propagating traveling wave
modes. Therefore, a two mode model with amplitudes $E_{1}$ and $E_{2}$ is
sufficient and the modes with amplitudes $E_{3}$ and $E_{4}$ can be dropped.
Similar to the equations (\ref{eq:statfields}) we look at the amplitudes after
one round trip time $\tau=1/\nu_{fsr}$.%

\begin{align}
E_{1}\left(  t+\tau\right)   &  =tr_{c}E_{1}\left(  t\right)  +r_{c}%
re^{-2ikx}E_{2}\left(  t\right)  +t_{in}E_{in,1}e^{i\varphi_{1}}%
\nonumber\label{eq:statfields2}\\
E_{2}\left(  t+\tau\right)   &  =tr_{c}E_{2}\left(  t\right)  +r_{c}%
re^{2ikx}E_{1}\left(  t\right)  ~.\nonumber\\
&
\end{align}
For simplicity, we drop the indices in $r_{1}$, $t_{1}$, and $x_{1}$ for a
single membrane. The equations of motion are obtained by first order
Taylor-expansion $E_{i}\left(  t+\tau\right)  \simeq E_{i}\left(  t\right)
+\tau\dot{E}_{i}\left(  t\right)  $. The result can be expressed as
\begin{equation}
\frac{\text{d}}{\text{d}t}\vec{E}=M(t)\vec{E}+\vec{\eta},\label{eq:dE_dt}%
\end{equation}
with matrix
\begin{align}
M &  =\nu_{fsr}\left(
\begin{array}
[c]{cc}%
A & B\\
C & A
\end{array}
\right)  ~,\nonumber\label{eq:M_1mem}\\
A &  :=tr_{c}-1~,\;\;B:=r_{c}re^{-i2kx}~,\;\;C:=r_{c}re^{i2kx}~,\nonumber\\
&
\end{align}
and field pump rate
\begin{align}
\vec{\eta} &  =\left(
\begin{array}
[c]{c}%
\eta\\
0
\end{array}
\right)  ,\;\eta:=\sqrt{\frac{T_{in}P_{in}}{\hbar kc}\nu_{fsr}}~.\nonumber\\
&
\end{align}
The matrix $M(t)$ contains the motion of the membrane $x(t)$ which varies
slowly in time as compared to the time scale of the cavity field dynamics. We
thus can solve equation (\ref{eq:dE_dt}) with the perturbation ansatz%
\begin{equation}
\vec{E}=\vec{E}_{a}\left(  t\right)  +\vec{\delta}\left(  t\right)
,\label{eq:ansatz}%
\end{equation}
with $\vec{\delta}\left(  t\right)  $ being the non adiabatic correction to
the instantaneous adiabatic solution $\vec{E}_{a}\left(  t\right)  =-M\left(
t\right)  ^{-1}\vec{\eta}$. Inserting the ansatz (\ref{eq:ansatz}) into
(\ref{eq:dE_dt}) provides a differential equation for $\vec{\delta}\left(
t\right)  $.
\begin{equation}
\frac{\text{d}}{\text{d}t}\vec{\delta}=M\left(  t\right)  \vec{\delta}%
+\frac{\text{d}}{\text{d}t}\left(  M\left(  t\right)  ^{-1}\vec{\eta}\right)
~.
\end{equation}
We approximately solve this equation by assuming the nonadiabatic correction
to be always in equilibrium, $d\vec{\delta}/dt=0$. This results in
\begin{equation}
\vec{\delta}\left(  t\right)  =-M\left(  t\right)  ^{-1}\frac{\text{d}%
}{\text{d}t}\left(  M\left(  t\right)  ^{-1}\vec{\eta}\right)
.\label{eq:delta_sol}%
\end{equation}
Inserting the matrix $M$ from Eq.~(\ref{eq:M_1mem}) into
Eq.~(\ref{eq:delta_sol}) the approximate solution of the problem
(\ref{eq:dE_dt}) can be written analytically as
\begin{align}
\vec{E}_{a} &  =-\frac{\eta\nu_{fsr}}{\det\left(  M\right)  }\left(
\begin{array}
[c]{c}%
A\\
-C
\end{array}
\right)  ~,\nonumber\\
\vec{\delta} &  =2ik\dot{x}\frac{\eta\nu_{fsr}^{2}}{\left(  \det\left(
M\right)  \right)  ^{2}}C\left(
\begin{array}
[c]{c}%
-B\\
A
\end{array}
\right)  ~,\nonumber\\
&
\end{align}
More formally speaking, this procedure is equivalent to replacing the total
differential $\frac{d}{dt}$ by $\frac{\partial}{\partial t}+v\frac{\partial
}{\partial x}$and expanding the fields in powers of the membrane velocity
$v=\dot{x}$, $E_{\pm}=E_{\pm}^{(0)}+vE_{\pm}^{(1)}+\mathcal{O}(v^{2})$
\cite{Domokos03}. Keeping only first order terms is a good approximation if
the energy decay rate of the cavity $\kappa$ is much larger than the
oscillation frequency of the membrane $\Omega_{m}$ (''Doppler limit''). For
our experiment this is well fulfilled ( $\Omega_{m}=2\pi\times130~\mathrm{kHz}%
$, $\kappa=2\pi\times3.6~\mathrm{MHz}$).

\subsection{Forces on the membrane}

The forces on the membrane are derived from momentum conservation for the
fields $\vec{E}_{1}$ and $\vec{E}_{2}$ and the membrane. Analogous to optical
forces on cold atoms, the forces can be separated into reflection and dipole
force
\begin{align}
F_{ref}  &  =2\hbar k\nu_{fsr}\cdot\left|  r\right|  ^{2}\left(  \left|
E_{1}\right|  ^{2}-\left|  E_{2}\right|  ^{2}\right) \nonumber\\
F_{dip}  &  =2i\hbar k\nu_{fsr}\operatorname{Im}\left(  r\right)  \left(
E_{1}E_{2}^{\ast}e^{2ikx}-E_{1}^{\ast}E_{2}e^{-2ikx}\right)  .\nonumber\\
&  \label{eq:forces}%
\end{align}
The radiation pressure force due to photon absorption in the membrane is very
small and can be neglected. By inserting the ansatz (\ref{eq:ansatz}) into
(\ref{eq:forces}) and in the approximation of small oscillation amplitude,
$x\ll\lambda$, the viscous parts of the forces proportional to $\delta
\propto\dot{x}$ take the form
\begin{align}
F_{ref,\dot{x}}  &  =4\hbar k\nu_{fsr}\cdot\left|  r\right|  ^{2}%
\operatorname{Re}\left(  E_{a,1}\delta_{1}^{\ast}-E_{a,2}\delta_{2}^{\ast
}\right)  ~,\nonumber\\
F_{dip,\dot{x}}  &  =4i\hbar k\nu_{fsr}\operatorname{Im}\left(  r\right)
\operatorname{Im}\left(  E_{a,1}\delta_{2}^{\ast}+E_{a,2}^{\ast}\delta
_{1}\right)  ~.\nonumber\\
&  \label{eq:forces1}%
\end{align}
The total damping force is related to the optomechanical damping rate by
$F=F_{ref,\dot{x}}+F_{dip,\dot{x}}=m_{eff}\Gamma_{opt}\dot{x}$ resulting in
\begin{align}
\Gamma_{opt}  &  =\Gamma_{0}\cdot K_{1}\cdot K_{2}~,\;\nonumber\\
\Gamma_{0}:=  &  \frac{T_{in}P_{in}}{m_{eff}c^{2}}kl~,\nonumber\\
K_{1}:=  &  \frac{8}{\left|  p^{2}-q^{2}\right|  ^{4}}~,\;\nonumber\\
K_{2}:=  &  \operatorname{Re}\left(  q^{\ast}\left(  p^{2}-q^{2}\right)
\right)  \times\nonumber\\
&  \times\left(  2\left|  r\right|  ^{2}\operatorname{Im}\left(  pq^{\ast
}\right)  -\operatorname{Im}\left(  r\right)  \left(  \left|  p\right|
^{2}-\left|  q\right|  ^{2}\right)  \right)  ~,\nonumber\\
p  &  :=tr_{c}-1\;\;\;\;q:=rr_{c}~.\nonumber\\
&
\end{align}

We are interested in the damping rates near the optical resonances of the
cavity. The frequency dependence is described by the complex round trip
reflectivity $r_{c}=r_{l}e^{ikl}$ which contains the phase $\phi=kl$ that the
light field accumulates during one round-trip. By analyzing the resonance
denominator of $K_{1}$ one finds that there are two resonances at $\phi
=\phi_{\pm}$ with $e^{i\phi_{\pm}}=r\pm t$. We expand $\Gamma_{opt}\left(
\phi\right)  $ around the resonance $\phi_{+}$ by setting $\cos\left(
\phi_{+}-\phi\right)  \approx1-\frac{1}{2}\left(  \phi_{+}-\phi\right)  ^{2}$.
The frequency detuning $\Delta$ and the cavity energy decay rate $\kappa$ can
be expressed by $\Delta=\nu_{fsr}\left(  \phi_{+}-\phi\right)  $ and
$\kappa=2\nu_{fsr}\left(  1-r_{l}\right)  $. We also make the standard
approximation $1-r_{l}^{2}=\left(  1+r_{l}\right)  \left(  1-r_{l}\right)
\approx2\left(  1-r_{l}\right)  =\frac{\kappa}{\nu_{fsr}}=\frac{\pi}{F}$ which
is valid for small round-trip losses, i.e. $r_{l}\approx1$. Finally one
arrives at an analytic expression for the optomechanical damping rate of a
single membrane in an ideal ring resonator:
\begin{equation}
\Gamma_{opt}=\frac{1}{4}\Gamma_{0}\cdot\nu_{fsr}^{2}\frac{-4\Delta\kappa
}{\left(  \kappa^{2}/4+\Delta^{2}\right)  ^{2}}%
\end{equation}
The extreme values
\begin{equation}
\Gamma_{\max}=\mp\frac{3\sqrt{3}}{8}\Gamma_{0}\cdot\left(  \frac{F}{\pi
}\right)  ^{2}~ \label{eq:Gamma_max}%
\end{equation}
are obtained for the detuning $\Delta=\pm\frac{1}{6}\sqrt{3}\kappa$. With the
numbers of our experiment $m_{eff}=3.88\times10^{-11}~\mathrm{kg}$, $\nu
_{fsr}=c/l=3.75~\mathrm{GHz}$, $F=520$, $T_{in}=t_{in}^{2}=6.04\times10^{-3}$,
and $P_{in}=10~\mathrm{mW}$ the maximum damping rate amounts to $\Gamma_{\max
}=-0.2~\mathrm{s}^{-1}$. This value is approximately 3 orders of magnitude
smaller than what we observe experimentally. Thus, the damping mechanism
analyzed so far cannot be the main reason for our observation. Good agreement
can be reached, however, if mirror back scattering is included.

\subsection{Mirror back scattering}

Unavoidable scattering at imperfections of a mirror surface may couple the
circulating modes of a ring cavity \cite{Krenz07}. In our experiment we have
thus used best available mirrors and the finesse of the empty ring resonator
can reach values of up to 150000. Still, back scattering is clearly detectable
and must be taken into account. A complete description is quite involved since
every optical surface in the cavity may contribute. Depending on the position
of these surfaces the various contributions interfere with different phases.
These interferences are different for each longitudinal mode. Furthermore, the
Purcell effect enhances scattering into cavity modes of high finesse. We
circumvent these details and assume that all scatterers together act like an
additional reflecting element inside the cavity with an effective complex
field reflectivity and transmittivity $r_{2}$ and $t_{2}$. The system is then
equivalent to the two membrane system shown in Fig.~\ref{fig:setup}(a) and the
equations (\ref{eq:statfields}) can be used as a starting point for deriving
the equations of motion. The procedure is the same as shown in detail in the
preceding section for a single membrane, with the difference that four light
fields are involved, and no practical analytical solution can be found. In
particular, Eq.~(\ref{eq:dE_dt}) is still valid with the matrix $M$ and the
pumping vector $\vec{\eta}$ now being four-dimensional. The matrix $M$ reads
\begin{widetext}
\begin{eqnarray}
M &:&=\left(
\begin{array}
{cccc} -\frac{1}{\tau_{r}} & 0 & \frac{1}{\tau_{r}}t_{2}r_{l1}e^{ikl-i\varphi_{d}} &
\frac{1}{\tau_{r}}r_{2}r_{l1}e^{ikl-i\varphi_{d}-2ikx_{2}}\\
0 & -\frac{1}{\tau_{d}} & \frac{1}{\tau_{d}}r_{2}r_{l2}e^{i\varphi
_{d}+2ikx_{2}} & \frac{1}{\tau_{d}}t_{2}r_{l2}e^{i\varphi_{d}}\\
\frac{1}{\tau_{d}}t_{1}r_{l2}e^{i\varphi_{d}} & \frac{1}{\tau_{d}}r_{1}%
r_{l2}e^{i\varphi_{d}-2ikx_{1}} & -\frac{1}{\tau_{d}} & 0\\
\frac{1}{\tau_{r}}r_{1}r_{l1}e^{ikl-i\varphi_{d}+2ikx_{1}} & \frac{1}{\tau
_{r}}t_{1}r_{l1}e^{ikl-i\varphi_{d}} & 0 & -\frac{1}{\tau_{r}}%
\end{array}
\right)  ~, \vec{\eta}=\left(
\begin{array}
{c} \eta\\  0\\0\\0
\end{array}
\right)  ~,
\end{eqnarray}
\end{widetext}
where we have introduced two different delay times $\tau_{d}:=\frac{d}{c}$ and
$\tau_{r}:=\frac{l-d}{c}$. The losses are distributed within the cavity:
$r_{l1}$ describes the losses along the path from one membrane to the other
membrane via the input coupler and $r_{l2}$ describes the losses along the
direct path between the membrane. The first membrane may oscillate while the
second membrane is held fixed at $x_{2}=0$. The stationary solution for the
nonadiabatic correction (\ref{eq:delta_sol}) now reads
\begin{equation}
\vec{\delta}\left(  t\right)  =2ik\dot{x}_{1}\left(  M^{-1}(t)\right)
^{2}DM^{-1}\vec{\eta},\label{eq:delta_sol3}%
\end{equation}
with
\begin{equation}
D=\left(
\begin{array}
[c]{cccc}%
0 & 0 & 0 & 0\\
0 & 0 & 0 & 0\\
0 & -\frac{1}{\tau_{d}}r_{1}r_{l2}e^{i\varphi_{d}} & 0 & 0\\
\frac{1}{\tau_{r}}r_{1}r_{l1}e^{ikl-i\varphi_{d}} & 0 & 0 & 0
\end{array}
\right)  ~.\label{eq:D_matrix}%
\end{equation}
\begin{figure}[th]
\centerline{\scalebox{.7}{\includegraphics{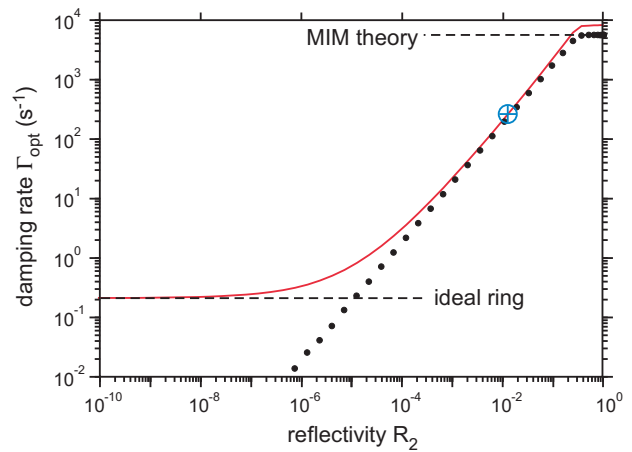}}}\caption{Optomechanical
damping rate of a membrane inside a ring cavity as function of the
reflectivity $R_{2}$ of an additional reflector. The separation between the
membrane and the reflector is set to $d=l/2$ and the losses are equally
distributed, $r_{1l}=r_{2l}$ with the total losses according to a Finesse of
$520$. In the standing wave limit, $R_{2}=1$, this choice allows a direct
comparison with the standard description of a MIM-Sytem according to Eq.
(\ref{eq:Gamma_lin_max}) (upper dashed line). The lower dashed line indicates the
result for a single membrane in an ideal ring cavity, Eq. (\ref{eq:Gamma_max}).
The black dots are the result of the expression Eq.~(\ref{eq:Gamma_lin}) by
inserting $G$ as determined from calculated ring cavity spectra. The blue
cross corresponds to the highest experimentally observed damping rate.}%
\label{fig:theory}%
\end{figure}
The solution for the instantaneous field $\vec{E}_{a}$ and the
correction $\vec{\delta}$ can easily be calculated numerically together with
the forces according to Eq.~(\ref{eq:forces1}). The calculations are done with
a separation between the membrane and the reflector set to $d=l/2$. The losses
are equally distributed, $r_{1l}=r_{2l}$ with total losses according to a
Finesse of $520$. The red line in Fig.~\ref{fig:theory} presents the resulting
maximum damping rates for various reflectivities $R_{2}=|r_{2}|^{2}$. For
decreasing $R_{2}$, mirror backscattering becomes less and less important. The
damping rate levels off and approaches the analytical solution for an ideal
resonator with a single membrane, (Eq.~(\ref{eq:Gamma_max}), lower dashed
line). In the opposite limit of $R_{2}=1$ the ring cavity is equivalent to the
standard geometry of a linear cavity with a membrane between the mirrors
(MIM). The damping rate of this limit is already reached for values of
$R_{2}>R_{1}=0.3$, above which the standing wave inside the resonator is
locked to the position of the second membrane. 

For MIM-systems an analytic expression for the damping rate in the Doppler
limit ($\kappa\gg\Omega_{m}$) is given in \cite{Aspelmeyer14}
\begin{equation}
\Gamma_{MIM}=-\frac{1}{2}G^{2}\frac{P_{in}}{m_{eff}\omega}\cdot T_{in}%
\nu_{fsr}\cdot\frac{4\kappa\Delta}{\left(  \kappa^{2}/4+\Delta^{2}\right)
^{3}}~.\label{eq:Gamma_lin}%
\end{equation}
The coefficient $G=$d$\omega_{c}/$d$x_{1}$ is the derivative of the cavity
resonance frequency
\begin{align}
\omega_{c} &  =\frac{c}{l_{MIM}}\arccos\left(  \sqrt{R}\cos\left(  2k\left(
x_{0}+x_{1}\right)  \right)  \right)  ~,\nonumber\label{eq:G}\\
&
\end{align}
with respect to the excursion $x_{1}$ of the membrane at the steady state
position $x_{0}$ \cite{Thompson08}. $R=\left|  r\right|  ^{2}$ is the
reflectivity of the membrane and $l_{MIM}$ is the mirror separation of the
standing wave resonator. Maximum damping $\Gamma_{MIM,\max}$ is obtained for
$2kx_{0}=\frac{\pi}{2}$ and $\Delta=\pm\frac{\sqrt{5}}{10}\kappa$. Using the
same notation as in Eq. (\ref{eq:Gamma_max}) one obtains
\begin{equation}
\Gamma_{MIM,\max}=-\frac{16\cdot200}{27}\sqrt{5}R_{1}\Gamma_{0}\left(
\frac{F}{2\pi}\right)  ^{4}~.\label{eq:Gamma_lin_max}%
\end{equation}
Note that for $R_{2}=1$ the resulting standing wave cavity has a mirror
separation of $2l$. With the parameters of our experiment the maximum damping
rate amounts to $\Gamma_{MIM,\max}=5.6\times10^{3}~\mathrm{s}^{-1}$, (upper
dashed line in Fig.~\ref{fig:theory}). This is slightly below the value
obtained from our four mode model. This discrepancy is to be expected since
Eq.~(\ref{eq:Gamma_lin}) is derived from a single mode model that does not
take into account the nonadiabatic correction to the mode function during the
oscillation of the membrane. Instead, the instantaneous steady state mode is
used and the model explains damping by the nonadiabatic correction to the mode
occupation alone. A full two mode model is described in ref. \cite{Jayich08}
but an analytic expression is not reported.

Our observed damping rate of $\Gamma\sim250~\mathrm{s}^{-1}$ (blue cross in
Fig.~\ref{fig:theory}) can now be explained if we assume an realistic value
for the effective mirror back scattering of $R_{2}\sim1\%$.

\subsection{Phase coupling vs. frequency coupling}

The damping mechanism in an ideal ring resonator in the limit of
$R_{2}\rightarrow0$ is fundamentally different from that usually discussed in
optomechanics. In the latter, a displacement of the membrane detunes the
cavity resonance relative to the frequency of the incoupled light which
changes the steady state light power in the cavity (''frequency coupling'').
In contrast, a membrane displacement in an ideal ring cavity has no effect on
the cavity resonance frequency and the steady state light power does not
change. In an ideal ring resonator it's the steady state position of the
standing wave inside the resonator that is shifted by the membrane (''phase
coupling''). In a real ring resonator with mirror back scattering, both
effects play a role. In order to distinguish the contributions of frequency
and phase coupling, we calculate the steady state resonance frequency for
various membrane positions $x_{0}$ and derive the frequency shift per membrane
excursion $G=$d$\omega_{c}/$d$x_{1}$ for each $x_{0}$. Similar as in linear
cavities one obtains a periodic function $G\left(  x_{0}\right)  $. We extract
the maximum value $G_{\max}=\max\left(  G(x_{0})\right)  $ and use Eq.~(\ref{eq:Gamma_lin}) to calculate the optomechanical damping rate. Its
dependency on the reflectivity $R_{2}$ is shown in Fig.~\ref{fig:theory} as
black dots. In the linear cavity limit ($R_{2}=1$), there is exact
agreement with the analytical single mode MIM-model (upper dashed line), as
expected. However, in the ring cavity limit, $R_{2}\rightarrow0$, the damping
rate tends to zero due to a vanishing value of $G$. This is in contrast with
the exact solution (red line) which levels off for values of $R_{2}%
\lesssim10^{-5}$. It is quite surprising that frequency coupling dominates
phase coupling already at such minute values for $R_{2}$. Pure phase coupling
thus seems hard to achieve experimentally.\\

\section{Summary and outlook}

We have analyzed the damping of a nanomembrane inside an optical ring cavity
and found a novel optical damping mechanism that is based on the coupling of
the membrane to the phase of the light field. Furthermore, we have identified
the reflectivity $R_{2}$ of a second reflector as crucial parameter for
modeling our experimentally observed damping rates. This second reflector is
used to describe mirror back scattering but it also may represent a second
membrane in the cavity. Since pure phase coupled damping ($R_{2}=0$) is small
and hard to observe it is probably more interesting for future work to
investigate ring resonators with two or more mechanical oscillators. We have
analyzed synchronization by applying the above described perturbation method
to a system of two moving membranes. For the parameters of our experiment we
find that the relative motion is damped with a high rate of several
$10^{4}~\mathrm{s}^{-1}$ which is comparable to the values obtained for
standard MIM-systems. This indicates that optical coupling of the oscillators
via the cavity fields leads to efficient phase synchronization. Ring cavities
thus seem to provide a promising playground for studying light mediated
coupling of two or more mechanical oscillators. If applied to oscillators in
their quantum ground state, such long-range coupling may allow for the
construction of entangled states and quantum state transfer between distant oscillators.

S. Slama is indebted to the Baden-W\"{u}rttemberg Stiftung for the financial
support of this research project by the Eliteprogramm for Postdocs. A. Yilmaz
is funded by Deutsche Telekom Stiftung. We acknowledge helpful discussion with
Florian Marquardt. Competing financial interests do not exist.


\begin{thebibliography}{9}
\bibitem {Sawadsky15_footnote}Please note that membranes in ring-like
Michelson-Sagnac interferometers like in \cite{Sawadsky15} are equivalent to
the membrane-in-the-middle setup.
\end{thebibliography}
\end{document}